\documentclass[superscriptaddress,twocolumn,10pt,aps,prb,floatfix,]{revtex4-1}

\usepackage{graphicx}
\usepackage{bm}
\usepackage{amsmath}
\usepackage{hyperref} \hypersetup{colorlinks=true,citecolor=blue,linkcolor=blue,urlcolor=blue}
\usepackage{amsmath}
\usepackage{multirow}
\usepackage{booktabs}
\usepackage{epsfig}
\usepackage{xcolor,soul}
\usepackage{epstopdf}
\usepackage{enumitem}

\makeatletter \renewcommand\frontmatter@abstractwidth{\dimexpr\textwidth\relax} \makeatother 

\setstcolor{red}    
\bibpunct{[}{]}{,}{n}{}{}   

\begin{document}

\def\AFLOW{{\small AFLOW}}
\def\ICSD{{\small ICSD}}
\def\citeAFLOW{\cite{aflowPAPER,aflowBZ,curtarolo:art110,curtarolo:art63,curtarolo:art57,curtarolo:art49,monsterPGM,curtarolo:art121}}
\def\citeAFLOWLIB{\cite{aflowlibPAPER,curtarolo:art92,curtarolo:art104,curtarolo:art128}}
\title{The structure and composition statistics of 6A \\ binary and ternary structures}

\author{Alon Hever}
\affiliation{Department of Physical Electronics, Tel-Aviv University, Tel-Aviv 69978, Israel}
\author{Corey Oses}
\affiliation{Department of Mechanical Engineering and Materials Science, Duke University, Durham, North Carolina 27708, United States}
\affiliation{Center for Materials Genomics, Duke University, Durham, North Carolina 27708, United States}
\author{Stefano Curtarolo}
\affiliation{Department of Mechanical Engineering and Materials Science, Duke University, Durham, North Carolina 27708, United States}
\affiliation{Center for Materials Genomics, Duke University, Durham, North Carolina 27708, United States}
\affiliation{Fritz-Haber-Institut der Max-Planck-Gesellschaft, 14195 Berlin-Dahlem, Germany}
\author{Ohad Levy}
\affiliation{Department of Mechanical Engineering and Materials Science, Duke University, Durham, North Carolina 27708, United States}
\affiliation{Center for Materials Genomics, Duke University, Durham,
  North Carolina 27708, United States}
\affiliation{Department of Physics, NRCN, P.O.Box 9001, Beer-Sheva 84190, Israel}
\author{Amir Natan}
\email{amirnatan@post.tau.ac.il}
\affiliation{Department of Physical Electronics, Tel-Aviv University, Tel-Aviv 69978, Israel}
\affiliation{Center for Materials Genomics, Duke University, Durham, North Carolina 27708, United States}
\affiliation{The Sackler Center for Computational Molecular and Materials Science, Tel-Aviv University, Tel-Aviv 69978, Israel}

\date{\today}

\begin{abstract}
The fundamental principles underlying the arrangement of elements into solid compounds with an enormous 
variety of crystal structures are still largely unknown. 
This study presents a general overview of the structure types appearing in an important 
subset of the solid compounds, i.e., binary and ternary compounds of the 6A column 
oxides, sulfides and selenides. 
It contains an analysis of these compounds, including the prevalence of various structure types, 
their symmetry properties, compositions, stoichiometries and unit cell sizes. It is found that these compound families include preferred stoichiometries and structure
types that may reflect both their specific chemistry and research bias in the
available empirical data. Identification of non-overlapping gaps and missing stoichiometries in these 
structure populations may be used as guidance in the search for new materials.

\end{abstract}

\maketitle

\section{Introduction}
The creation of novel materials with optimal properties for diverse applications requires a fundamental 
understanding of the factors that govern the formation of crystalline 
solids from various mixtures of elements.
Compounds of the non-metallic elements of column 6A, oxygen, sulfur and selenium, are of particular interest. 
They serve in a large variety of applications
in diverse fields of technology, e.g., chemistry, catalysis, optics, 
gas sensors, electronics, thermoelectrics, piezoelectrics, 
topological insulators, spintronics and more~\cite{eranna2004oxide,fortunato2012oxide,tsipis2008electrode,jiang1998new,panda2009review,shi2017,lorenz20162016,ruhle2012all}. 
Given the very large number of possibilities, many of the alloy systems of these elements have not 
been fully investigated, some of them even not at all.

In recent years, high-throughput computational techniques based on {\it ab-initio} calculations 
have emerged as a potential route to bridge these experimental gaps and
gain understanding of the governing principles of compound formation~\cite{nmatHT}.
This led to the creation of large databases of computational materials
data \cite{curtarolo:art65,CMS_Ong2012b}.
Yet, these computational approaches are practically limited by the number and size of structures 
that can be thoroughly analyzed, and fundamental issues that limit 
the applicability of standard semi-local DFT for non-metallic compounds. 
The sought-after governing principles are thus still largely unknown. 

Nevertheless, the considerable body of experimental data that is already available, 
although by no means complete, is a useful basis for large-scale data analysis. 
This experimental data is usually presented in
compendiums that lack statistical analysis.
Presenting this data in a structured manner may be conducive for gaining insights 
into the essential factors that determine structure formation, and may help to provide 
material scientists with the necessary foundation for rational
materials design.

Analyses recently carried out for the intermetallic binaries~\cite{dshemuchadse2014some} 
and ternaries~\cite{dshemuchadse2015more} have uncovered interesting Bravais lattices distributions and an unexpected large prevalence of unique structure types. 
Here we extend the analysis and discuss trends, as well as special phenomena, across 
binary and ternary compounds of the 6A non-metals.
This analysis reveals the following
interesting observations:
\begin{itemize}[leftmargin=*]

  \item Considerable overlap exists between the sulfides and selenides:
   about a third of the total number of structure types are shared among 
  both compound families. 
  In contrast, the overlap between the oxides and the other two families is rather small.
	
  \item The prevalence of different compound stoichiometries in the sulfide
	and selenide families is very similar to each other
	but different from that of the oxides. Some stoichiometries
	are abundant in the oxides but are {\it almost
  absent} in the sulfides or  selenides, and vice versa. 
  
  \item The number of ternary oxide stoichiometries, $A_{x}B_{y}$O$_{z}$, decreases when the product of 
    binary oxide stoichiometries, of participating elements, increases. This behavior can be explained by general thermodynamic arguments and is discussed in the text.
    
  \item Overall, oxide compounds tend to have richer oxygen content than the sulfur and selenium content in their corresponding compounds.  
  
  \item Across all three compound families, most structure types are represented 
    by only one compound.
 
  \item High symmetry lattices, e.g.\ the orthorhombic face centered,
    orthorhombic body centered and cubic lattices
	  are relatively rare among these compounds. 
    This reflects the spatial arrangement of the compound forming orbitals of the 6A non-metals, 
    whose chemistry does not favor these structures.

\end{itemize}

In the analysis presented here, we adopt the ordering of the elements by Mendeleev numbers as 
defined by Pettifor \cite{pettifor:1984,pettifor:1986}, 
and complement it by investigating the crystallographic properties of
the experimentally reported compounds. 
Pettifor maps constructed for these compound families exhibit similar separation between different structure types as the
classical Pettifor maps for binary structure types \cite{pettifor:1984,pettifor:1986}.
For some stoichiometries, the structure types show similar patterns in
the maps of the three compound families, suggesting
that similar atoms tend to form these stoichiometries with all three elements. 
Such similarity of patterns is more common between
the sulfides and selenides than between either of them and the oxides. 

These findings suggest a few possible guiding principles for directed searches of new compounds. 
Element substitution could be used to examine favorable candidates within the 
imperfect overlaps of the structure distributions, especially between the sulfides and selenides.
Moreover, the missing stoichiometries and structure symmetries mean that data-driven approaches, e.g.,
machine learning, must use training sets not limited to one compound family, even in studies  directed at that specific set of compounds.
This hurdle may be avoided by augmenting the known structures with those of the other families. 
In addition, identified gaps in the Mendeleev maps suggest potential new compounds, 
both within each family or by correlations of similar structure maps across the different families.

\section*{Data methodology}

\begin{table*} [t]
	
	\begin{tabular}{ |p{3.5cm}||p{3.5cm}|p{3.5cm}|p{3.5cm}|  }
		\hline
		 & compounds & unique compounds & structure types \\
		\hline \hline
		total                            & 88,373    & 50,294& 13,324   \\
		\hline
		unary              & 1752    & 499   & 197\\
		\hline \hline
		binary & 27,487 & 10,122 & 1,962 \\
		\hline
		binary oxides  & 3,256 & 844 & 538 \\ \hline
		binary sulfides & 1,685 & 495 & 270 \\ \hline
		binary selenides & 1,050 & 332 & 168 \\ \hline \hline
		ternary  & 37,907 & 23,398&  4,409\\
		\hline
		ternary oxides  & 10,350 & 5,435 & 2,079 \\ \hline
		ternary sulfides & 3,190 & 2,041 & 784 \\ \hline
		ternary selenides & 1,786 & 1,256 & 521\\ \hline \hline
		quaternary & 15,138 & 11,050 & 3,855 \\
		\hline
		5 atoms         & 4,638 & 3,899 & 2,053 \\
		\hline
		6 atoms & 1,219 & 1,101 & 682 \\
		\hline
		7 atoms & 212	& 201 &	154 \\
		\hline
		8 atoms          & 20  & 20   & 12\\
		\hline
		
	\end{tabular}
	\caption{Data extraction numerical summary.}
	\label{table:ICSD_DATA}
\end{table*}

The \ICSD~\cite{ICSD_database} includes approximately 169,800 entries (as of August 2016). 
For this study we exclude all entries with partial or random occupation and those that do not have full structure data. 
The remaining set of structures has been filtered using the \AFLOW\ software~\citeAFLOW, 
which uses an error checking protocol to ensure the integrity of each entry. 
\AFLOW\ generates each structure by appropriately propagating the Wyckoff positions of the specified spacegroup. 
Those structures that produce inconsistencies, e.g., overlapping atoms or a different stoichiometry 
than the structure label are ignored. 
If atoms are detected to be too close ($\leq 0.6$\AA), alternative standard ITC 
(International Table of Crystallography)~\cite{tables_crystallography} settings of the spacegroup are attempted. 
These settings define different choices for the cell's unique axes, possibly 
causing atoms to overlap if not reported correctly. 
Overall, these considerations reduce the full set of \ICSD\ entries to a
much smaller set of 88,373 ``true'' compounds. 
These entries are contained in \AFLOW\ Database~\citeAFLOWLIB.
{They include the results of the AFLOW generated full symmetry analysis for each structure, i.e. Bravais lattice, 
space group and point group classifications, and Pearson symbol 
(the method and tolerances used for this analysis follow the AFLOW standard \cite{curtarolo:art104}).} 
For the analysis presented here we identify all the binary and ternary compounds included in this set,
27,487 binary entries and 37,907 ternary entries.
From these, we extract all the entries that contain oxygen, sulfur or selenium as one of the components. 
Of the binaries, we find 3,256 oxides, 1,685 sulfides and 1,050 selenides.
10,530 oxides, 3,190 sulfides and 1,786 selenides are found among the ternaries.
Duplicate entries representing different experimental reports of the same compound, 
i.e., the same elements, stoichiometry, space group and Pearson designation, are then eliminated 
to obtain a list in which every reported compound is represented by {its most recent corresponding entry in the ICSD}. 
This reduces our list of binaries to 844 oxides, 495 sulfides and 332 selenides, and 
the list of ternaries to 5,435 oxides, 2,041 sulfides and 1,256
selenides.
These results are summarized in Table~\ref{table:ICSD_DATA}. 
Throughout the rest of the paper, we will refer to these sets of
binary and ternary compounds. We choose not to discuss multi-component structures with four or more elements since their
relative scarcity in the database most probably indicates incomplete
experimental data rather than fundamental issues of their chemistry. 
{It is also instructive to check the effect of element abundance on the number of compounds. The abundance of oxygen in the earth's crust is $\sim$ 47$\%$ by weight, around 1000 times more than that of sulfur ($\sim 697$ppm) which is around $5,000$ more abundant than selenium ($120$ppb)\cite{wedepohl1995composition}. Comparison with the number of elements (O/S/Se) binary compounds, 844/495/332, or ternary compounds, 5,435/2,041/1,256, makes it clear that while a rough correlation exists between the elements abundance and the number of their known compounds, it is by no means a simple proportion.}
\begin{figure} [t]
	\centering
  \includegraphics[width=0.9\linewidth]{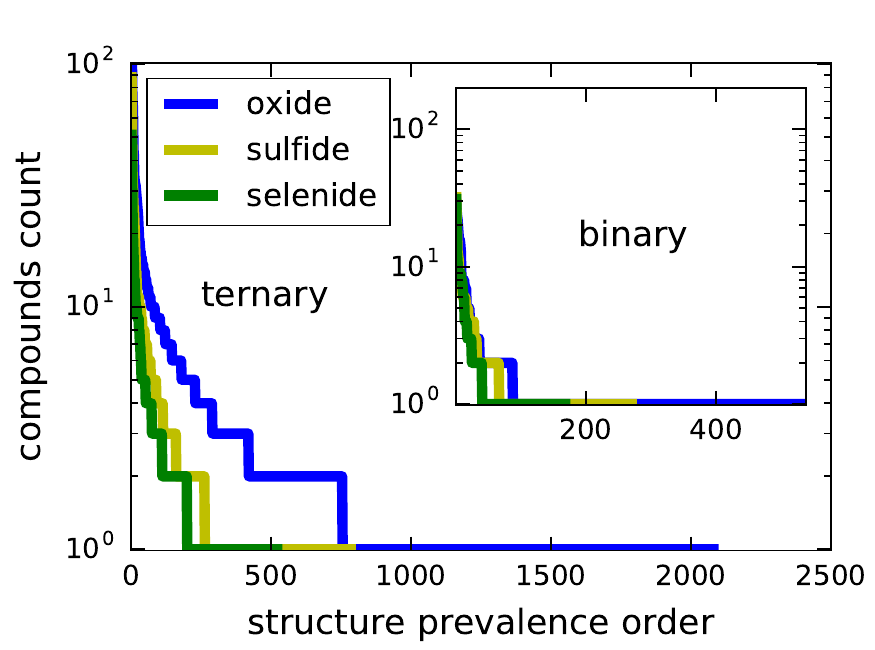}
  \caption{
    Distributions of the compounds among structure types for binary (inset) and ternary compounds.
    Oxides are shown in blue, sulfides in yellow and selenides in green. {The binary distributions differ mostly by the length of their single-compound prototypes tails, while the ternary distribution of the oxides deviates significantly from those of the sulfides and selenides.}
    }
	\label{fig:prototypes_distribution_curves_log}
\end{figure}

In the next stage, we identify unique structure types.
Structure types are distinguished by stoichiometry, space group, and Pearson designation, without consideration
of the specific elemental composition.
This implicit definition of structure type is common in the literature~\cite{Villars2013, PaulingFile},
and we use it throughout the manuscript as 
providing a good balance of clarity and simplicity. 
However, it should be noted that there are a few rare cases of complex structures where a given 
structure type under this definition includes a few sub-types (see Figure $S1$ in the Supporting Information). 
Examples exist of more complex definitions of structure types, formulated to define similarities 
between inorganic crystals structures \cite{lima1990nomenclature}. 

The binary structure type lists contain 538 oxides, 270 sulfides and 168 selenides.
The ternary lists contain 2,079 oxides, 784 sulfides and 521 selenides. 
This means that 64\% of the binary oxides, 55\% of the sulfides and 51\% of the selenides are distinct structure types. 
The corresponding ratios for the ternaries are 38\% of the oxides, 38\% of the sulfides and 41\% of the selenides.
All the other entries in the compound lists represent compounds of the same 
structure types populated by different elements.
Differently put, this means that there are on average about 1.6 compounds per structure type in the binary oxides, 
$1.8$ in the binary sulfides and $2$ in the binary selenides. 
Among the ternaries, the corresponding numbers are 2.6 compounds per structure type in the oxides, 
$2.6$ in the sulfides and $2.4$ in the selenides.
These numbers may be compared to the intermetalllics, where there are
20,829 compounds of which 2,166, about 10\%, 
are unique structure types \cite{dshemuchadse2014some}.
There are about seven compounds per structure types in the binary intermetallics and about nine in the ternaries. 
The number for binary intermetallics is considerably larger than for ternary oxides, sulfides or selenides. Together with the higher proportion of unique structure types in the latter, this reflects the limits on 
materials chemistry imposed by the presence of one of those 6A elements.

\begin{figure*} [t]
	\centering
	\includegraphics[width=1.0\linewidth]{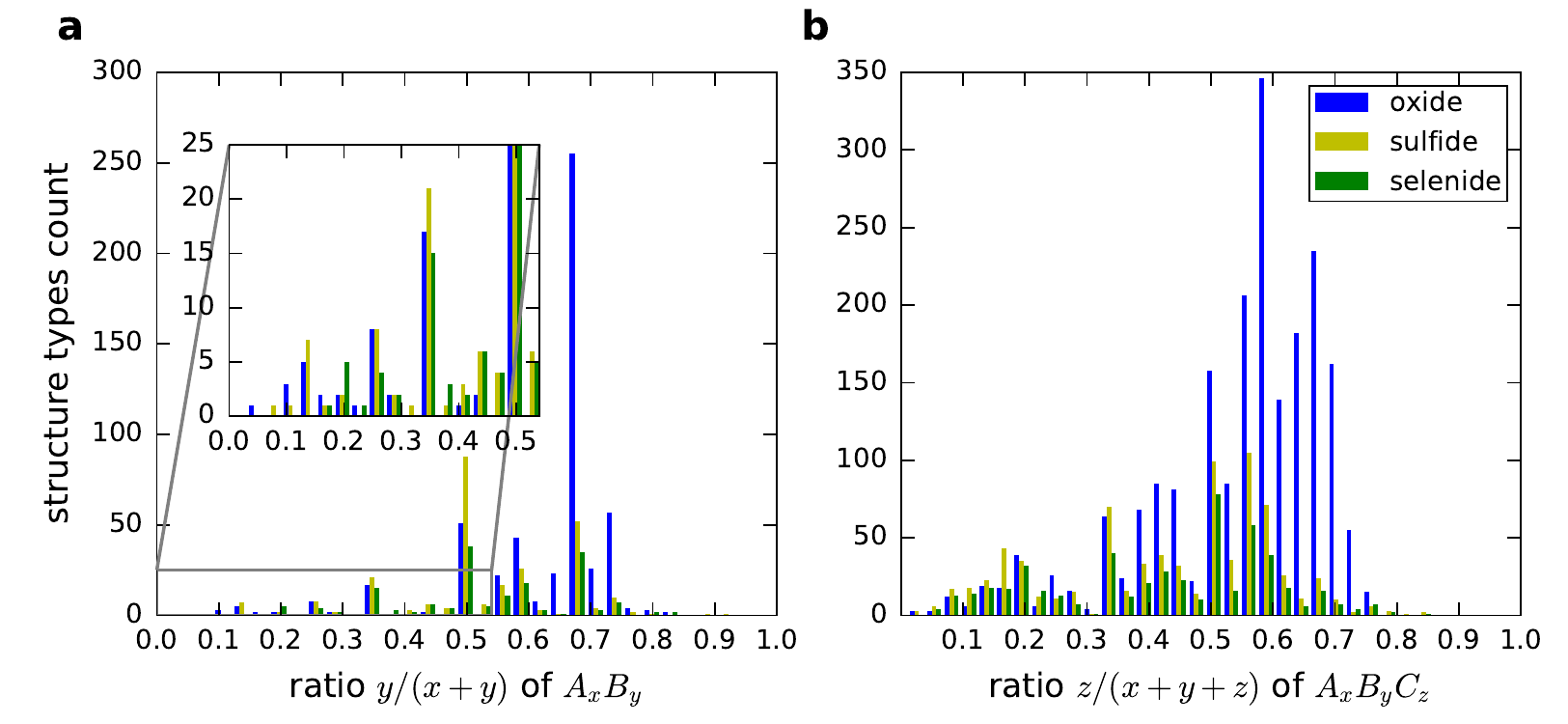}
  \caption{The distribution of (\textbf{a}) binary stoichiometries and (\textbf{b}) ternary stoichiometries.
		Oxides are shown in blue, sulfides in yellow and selenides in green. {The distributions of the selenides and sulfides are quite similar while those of the oxides deviate significantly, as detailed in the text.}
    }
	\label{fig:stoi_hist}
\end{figure*}

It should be noted that this structure selection procedure produces lists that partially overlap, 
i.e., certain structure types may appear in more than one list, 
since there might be oxide structure types that are also represented among 
the sulfide or selenide structures, and vice versa.
11\% of the oxide binary structure types also appear in the sulfides binary list and 8\% are represented 
in the selenides binary list. 
33\% of the sulfide binaries are also represented in the selenides list. 
The total number of binary oxides, sulfides and selenides structure types is 976, which is reduced by 16\%,
to 818 structure types, by removing all overlaps. 
The corresponding overlap ratios for the ternaries are 10\% for the oxides and sulfides, 
6\% for the oxides and selenides and 31\% for the sulfides and selenides. 
The total number of entries in the ternary oxides, sulfides, and selenides structure type lists is 3,384, 
which is reduced to 2,797 structure types by removing all overlaps, a 17\% reduction.
Therefore, the overlaps between these three compound families are similar for the binaries and ternaries. 
In both, the overlap between the oxides and the other two families is rather small, 
whereas the overlap between the sulfides and selenides represents about a third of the total number of structure types.

The sequence of Mendeleev numbers includes 103 elements, from hydrogen to lawrencium
with numbers 1-6 assigned to the noble gases, 2-16 to the alkali metals and alkaline earths, 
17-48 to the rare earths and actinides, 49-92 to the metals and metalloids and 93-103 to the non-metals.
Of these, noble gases are not present in compounds and artificial elements 
(metals heavier than uranium) have very few known compounds. 
We are thus left with 86 elements, of which the above compounds are composed.
That means there are about ten times more oxide binaries than 
element-oxygen combinations, about six times more sulfides than element-sulfur
combinations and four times more selenides than element-selenium combinations.
Oxides are much more common than sulfides and selenides.
The corresponding numbers for the ternaries are much lower. 
There are about 1.6 times more ternary oxides than two-element-oxygen ternary possible systems, 
about 0.6 times less ternary sulfides and about 0.4 times less ternary selenides than the corresponding two-element combinations.
The ternaries are relatively quite rare, more so as we progress from oxides to sulfides and then to selenides. 
A similar analysis of the intermetallic binaries in Reference~\onlinecite{dshemuchadse2014some} shows that of the 20,829 intermetallics, 
277 are unaries (about three times more than possible metal elements), 6,441 are binaries
(about two times more than possible metal binary systems),
and 13,026 are ternaries (6.5 times less than possible metal ternary systems).
This means that unary metal structures are less common among the metallic 
elements than the oxide, sulfide and selenide binary compounds among their corresponding binary systems.
This seems to reflect simply the larger space of stoichiometries available to binaries over unaries.
However, on the contrary, the intermetallic binary compounds are more common among the metallic binary 
systems than the oxide, sulfide and selenide ternary compounds among their corresponding ternary systems.
This discrepancy again reflects either the chemical constraints imposed by the presence of a 6A non-metal on the 
formation of a stable ternary structure, or simply gaps in the 
experimental data since many ternary systems have not been thoroughly investigated. 

\begin{figure*}
	\centering
  \includegraphics[width=1.0\textwidth,height=0.85\textheight,keepaspectratio]{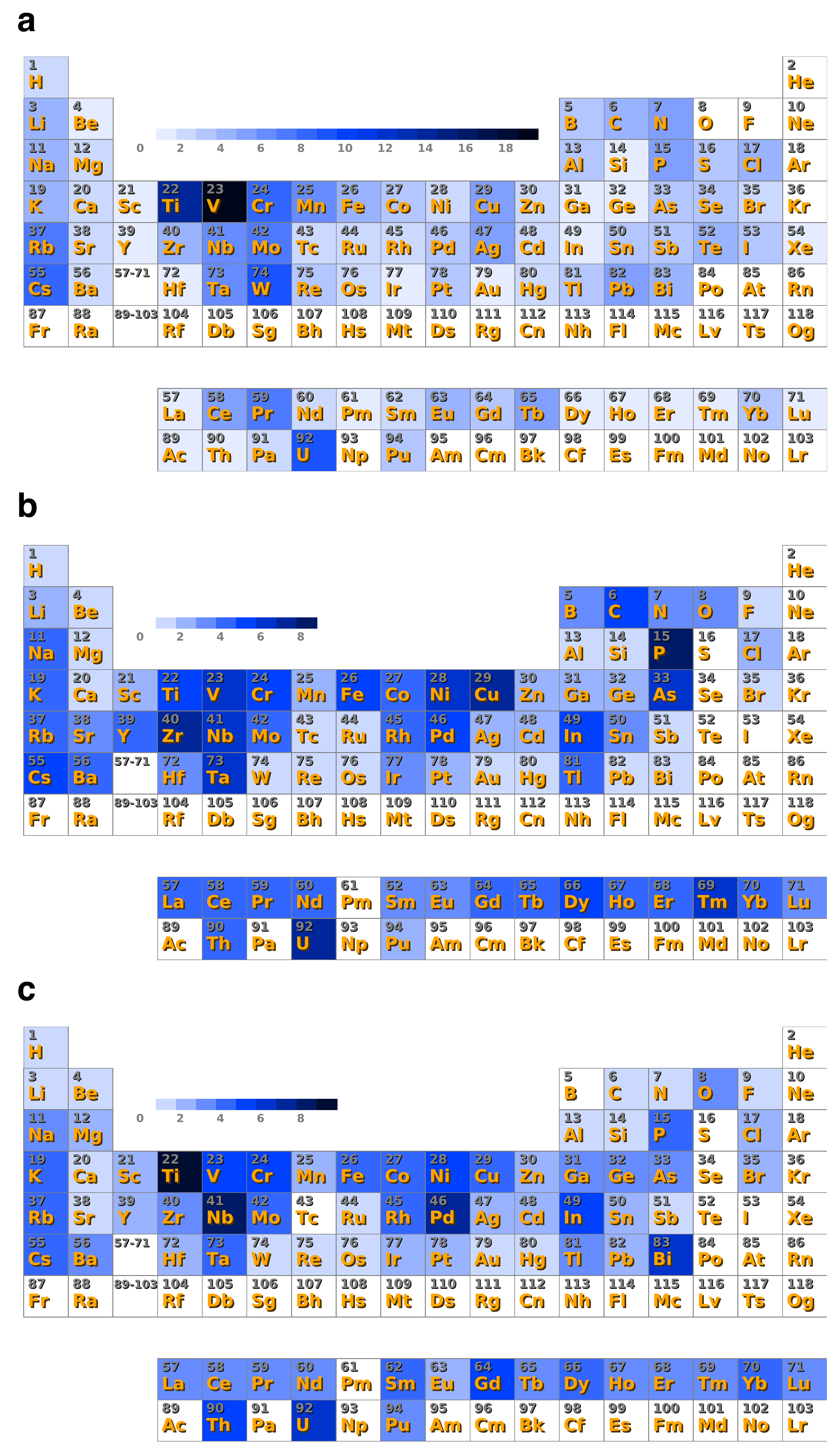}
  \caption{Binary (\textbf{a}) oxide, (\textbf{b}) sulfide, and (\textbf{c}) selenide stoichiometries {(number of different stoichiometries that include the respective element)
  	. The colors go from no stoichiometries (white) to the maximal number of stoichiometries (dark blue) which is different for each element, 19/8/9 for O/S/Se. Islands of high prevalence appear for the 4B and 5B transition metals and the heavy alkalies in all three compound families. Additional, smaller islands appear in the sulfides and selenides for the 8 and 1B transition metals and  the 3A and 5A semi-metals.}
	}
	\label{fig:stoi_periodic_oxide}
\end{figure*}

\section{Results and discussion}
\noindent \textbf{Structure types.}
The distribution of the binary and ternary compounds among the corresponding structure types is shown 
in Figure~\ref{fig:prototypes_distribution_curves_log}. 
Detailed data for the most common structure types is presented in Tables $S1-S6$ in the Supporting Information. 

About 84\% of the binary oxide structure types represent a single
compound, characterizing the tail end of the binary oxide distribution.
They include about 53\% of the binary oxide compounds.
The most common structure type represents 29 compounds,
3.4\% of the oxide compounds list.
Among the binary sulfides, 76\% of the structure types represent a single compound.
They include 41\% of the binary sulfide compounds.
The most common structure type represents 32 compounds, 6.5\% of the
sulfide compounds list. 
Among the binary selenides, 76\% of the structure types represent a single compound.
They include 39\% of the binary selenide compounds. 
The most common structure type represents 31 compounds, 9.3\% of the selenide compounds list. 

In all three binary lists the most common structure type is rock salt (NaCl). 
The binary oxide structure type distribution has a much longer tail than the sulfides and selenides, 
i.e., more oxide compounds have unique structure types. 
The most common structure type in these three distributions represents
a similar number of compounds but a smaller proportion of the corresponding compounds in the oxides. 
The middle regions of the distributions are very similar 
(inset Figure~\ref{fig:prototypes_distribution_curves_log}).
This means that the much larger number of binary oxide compounds, compared to the sulfides and selenides, 
is expressed at the margin of the distribution, in the long tail of unique compounds.

This discrepancy between the three binary distributions is much less
apparent among the ternary compounds. 
64\% of the ternary oxide structure types represent a single compound.
They include 24\% of the ternary oxide compounds. 
The two most common structure types, pyrochlore and perovskite, represent 116 and 115 compounds, 
respectively, about 2\% each of the entire compounds list. 
Among the ternary sulfides, 70\% of the structure types represent a single compound.
They include 34\% of the ternary sulfide compounds. 
The most common structure type, delafossite, represents 65 compounds,
4\% of the entire compounds list. 
Among the ternary selenides, 62\% of the structure types represent a single compound.
They include 26\% of the ternary selenide compounds. 
The most common structure type, again delafossite, represents 51 compounds, 4\% of the ternary sulfides. 

In contrast to the binaries, the larger count of ternary oxides, compared to the sulfides and selenides, 
is expressed by a thicker middle region of the structure type distribution, 
whereas the margins have a similar weight in the distributions of the three compound families.

\noindent \textbf{Binary stoichiometries.}
The structure types stoichiometry distribution for the binary oxide, sulfide and selenide compounds is shown in 
Figure~\ref{fig:stoi_hist}(a). 
We define the binaries as $A_xB_y$, where $B$ is O, S or Se, and the number of structure types is shown as a function of $y/(y+x)$. 
A very clear peak is found for the oxides at the stoichiometry 1:2, 
$A$O$_2$, while both the sulfides and selenides have a major peak at 1:1, $A$S and $A$Se, respectively. 

For $y/(y+x)<0.5$, there are more gaps in the plot (missing stoichiometries) for the oxides compared 
to the sulfides and selenides, while for $y/(y+x)>0.6$ there are more gaps in the sulfides and selenides, 
this behavior is shown in detail in Table $S10$ in the SI. 
An important practical conclusion is that augmenting the binary oxide structure types with 
those of sulfides and selenides will produce a more extensive coverage of possible stoichiometries.

Another interesting property is the number of stoichiometries
for each of the elements in the periodic table. 
The prevalence of binary oxide stoichiometries per element is shown in Figure \ref{fig:stoi_periodic_oxide}(a). 
A few interesting trends are evident --- the first row of transition metals shows a peak near vanadium (19 stoichiometries) 
and titanium (14 stoichiometries). 
Hafnium, which is in the same column of titanium has only a single stoichiometry --- HfO$_{2}$. 
Both the beginning and end of the $d$-elements exhibit a small amount of stoichiometries --- scandium 
with only one and zinc with only two. 
The two most abundant elements, silicon and oxygen, form only a single stoichiometry in the 
\ICSD\ --- SiO$_{2}$, with 185 {\it different} structure types. 
Another interesting trend is evident for the alkali metals, where rubidium and cesium have more 
stoichiometries --- perhaps related to the participation of $d$-electrons in the chemical bonds.

Figures \ref{fig:stoi_periodic_oxide}(b) and (c)
show the binary stoichiometries prevalence per element for sulfur and selenium respectively. 
Similar trends are exhibited --- there are two ``islands'' of large number of stoichiometries 
in the transition metals: one around vanadium and titanium and the other near nickel and copper. 
Evidently, prime candidates for new compounds should be searched among structures in the vicinity of 
these high density islands, especially for elements that exhibit a considerably higher density in one family.

\noindent \textbf{Ternary stoichiometries.} 
Similar to the binaries, the ternary stoichiometries are designated
$A_xB_yC_z$, where $C$ is O, S or Se.
The distributions of the ternaries are, as might be expected, more complex, 
with maxima at $z/(x+y+z)=0.6$ for the oxides, $z/(y+x+z)=0.55$ for the sulfides and $z/(y+x+z)=0.5$ for the selenides. 
The major peaks still appear at integer and half integer values, but with more minor peaks at intermediate values.
This behavior is shown in Figure \ref{fig:stoi_hist}(b).
The ternary selenide and sulfides distributions are again nearly identical, and there are 
almost no compounds with ratios larger than $0.75$ in the oxides or larger than $0.66$ in the sulfides and selenides. 
However, there are few sulfide and selenide compounds around 0.8 and 0.85 but no oxides. 

\begin{table*} [t]
	\caption{Ternary stoichiometry data: $A_xB_yC_z$. ``$C$-rich'' 
		refers to stoichiometries where $z>x+y$.}
	\begin{tabular}{ |p{6cm}||p{2cm}|p{2cm}|p{2cm}|  }
		\hline
		& oxygen & sulfur & selenium \\
		\hline
		Number of stoichiometries    & 585   & 282 & 206   \\
		\hline
		$C$-rich stoichiometries ratio & 0.85    & 0.67   & 0.66 \\
		\hline 
		$C$-rich compound ratio  & 0.92 & 0.77 & 0.73 \\                             
		\hline
	\end{tabular} 
	\label{table:ternary_stoi}
\end{table*}

\begin{figure*}
	\centering
	\includegraphics[width=1.0\linewidth]{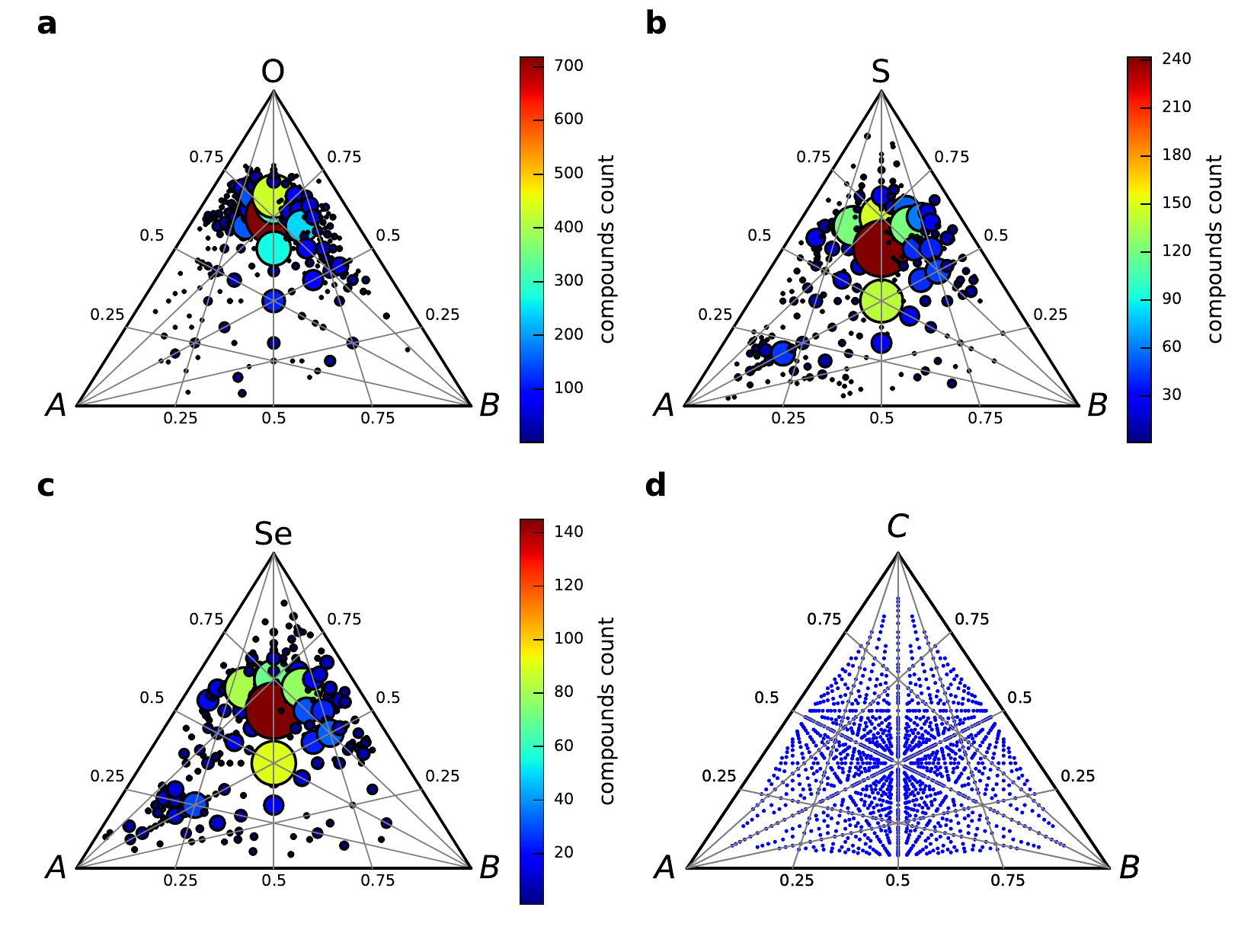}
	\caption{Prevalence of stoichiometries among the ternary (\textbf{a}) oxide, 
		(\textbf{b}) sulfide and (\textbf{c}) selenide compounds. 
		(\textbf{d}) shows, for reference, all the possible stoichiometries with up to 12 atoms
		of each component per unit cell. 
		In each figure, the smaller circles are normalized to the biggest one, which denotes the highest prevalence, i.e.,
		718 for oxides, 242 for sulfides, and 145 for selenides{, in addition a heat map color scheme is used where blue means low prevalence and red means the highest prevalence for each element}. 
		The $x$ and $y$ axes denote the atomic fractions in the ternaries $A_xB_yC_z$, where $C$ is O, S or Se, respectively. 
		$A$ and $B$ are ordered by Mendeleev number where $M_A>M_B$.
	}
	\label{fig:triangles}
\end{figure*}

Another perspective of ternary stoichiometries is demonstrated in Figure~\ref{fig:triangles} 
which shows the abundance of the most common stoichiometries.
The biggest circle in each diagram denotes the prevalence of the most common stoichiometry 
(number of unique compounds for this stoichiometry), 
which is 718 ($x=1$, $y=1$, $z=3$) for oxides, 242 ($x=1$, $y=1$, $z=2$) for sulfides, and 145 ($x=1$, $y=1$, $z=2$) for selenides. 
The smaller circles in each plot are normalized to the corresponding highest prevalence. 

These diagrams highlight the similarities as well as important differences between the three families of compounds. 
In all three cases, the most common stoichiometries appear on the symmetry axis of the diagram, i.e.,
at equal concentrations of the $A$ and $B$ components, or very close to it. 
For the oxides, they are concentrated near 0.5-0.6 fraction of oxygen, representing the 
$A_1B_1$O$_2$ and $A_1B_1$O$_3$ stoichiometries, respectively, 
and form a very dense cluster with many similar reported stoichiometries of lower prevalence.  
Outside this cluster, the occurrence of reported compositions drops sharply, and other regions 
of the diagram are very sparsely populated, in particular near the vertices of the $B$ and O components. 

The sulfide and selenide diagrams also exhibit prominent clusters on the $AB$ symmetry axes, 
but they appear at a lower S or Se concentration of about 0.5, i.e., $A_1B_1C_2$ stoichiometry. 
They are considerably more spread out and include a significant contribution at the $ABC$ stoichiometry. 
In both sulfides and selenides, an additional minor cluster appears closer to the $A$ vertex (Figure~\ref{fig:triangles}).
A few members of this cluster are ternary oxides, reflecting the high electronegativity and high Mendeleev number (101) of oxygen. 
The $B$ and $C$ vertex regions are still sparsely populated, but less so than in the oxides case. 
Overall, the sulfide and selenide diagrams are very similar to each other and different from that of the oxides. 
They are more spread out, less $AB$ symmetric than the oxide diagram and less tilted towards rich $C$-component concentration. 
This discrepancy may reflect some uniqueness of oxygen chemistry compared to sulfur and selenium, 
or rather simply reflect the oxygen rich environment in which naturally formed compounds are created in the atmosphere. 
The number of stoichiometries and the differences in the $C$-component concentration are summarized in Table~\ref{table:ternary_stoi}.  

\begin{table*} [t]
	\centering
	\caption{Distribution of the oxide, sulfide and selenide compounds and structure types among the 14 Bravais lattices.}
	\label{table:bravais_lattice_distribution}
	\resizebox{\linewidth}{!}{
		\begin{tabular}{|p{1cm}|p{1cm}|p{1cm}|p{1cm}|p{1cm}|p{1cm}|p{1cm}|p{1cm}|p{1cm}|p{1cm}|p{1cm}|p{1cm}|p{1cm}|p{1cm}|p{1cm}|p{1cm}|p{1cm}|p{1cm}|p{1cm}|}
			\hline
			& \multicolumn{3}{p{3cm}|}{binary \newline compounds} & \multicolumn{3}{p{3cm}|}{binary \newline structure types}
			& \multicolumn{3}{p{3cm}|}{binary \newline compounds per \newline structure type} 
			& \multicolumn{3}{p{3cm}|}{ternary \newline compounds} & \multicolumn{3}{p{3cm}|}{ternary \newline structure types}
			& \multicolumn{3}{p{3cm}|}{ternary \newline compounds per \newline structure type}\\ \hline
			& O & S & Se & O & S & Se & O & S & Se & O & S & Se & O & S & Se & O & S & Se \\ \hline
			aP & 51 & 13 & 5 & 39 & 12 & 5 & 1.3 & 1.1 & 1 & 378 & 79 & 60 & 219 & 56 & 39 & 1.7 & 1.4 & 1.5 \\ \hline
			mP & 82 & 54 & 31 & 62 & 36 & 20 & 1.3 & 1.5 & 1.6 & 918 & 318 & 198 & 363 & 166 & 109 & 2.5 & 1.9 & 1.8 \\ \hline
			mS & 88 & 31 & 22 & 58 & 21 & 15 & 1.5 & 1.5 & 1.5 & 672 & 251 & 170 & 292 & 117 & 77 & 2.3 & 2.1 & 2.2 \\ \hline
			oP & 123 & 82 & 48 & 81 & 37 & 30 & 1.5 & 2.2 & 1.6 & 950 & 481 & 266 & 373 & 139 & 105 & 2.5 & 3.5 & 2.5 \\ \hline
			oS & 39 & 24 & 11 & 36 & 19 & 9 & 1.1 & 1.3 & 1.2 & 334 & 84 & 60 & 133 & 40 & 25 & 2.5 & 2.1 & 2.4 \\ \hline
			oF & 11 & 7 & 11 & 10 & 6 & 4 & 1.1 & 1.2 & 2.8 & 51 & 32 & 23 & 28 & 14 & 8 & 1.8 & 2.3 & 2.9 \\ \hline
			oI & 22 & 5 & 2 & 20 & 4 & 2 & 1.1 & 1.25 & 1 & 89 & 36 & 27 & 39 & 15 & 12 & 2.3 & 2.4 & 2.25 \\ \hline
			tI & 41 & 20 & 10 & 31 & 17 & 8 & 1.3 & 1.2 & 1.25 & 418 & 80 & 72 & 101 & 34 & 23 & 4.1 & 2.4 & 3.1 \\ \hline
			tP & 78 & 27 & 28 & 48 & 13 & 16 & 1.6 & 2.1 & 1.75 & 239 & 73 & 52 & 107 & 39 & 26 & 2.2 & 1.9 & 2.0 \\ \hline
			hP & 94 & 87 & 66 & 62 & 50 & 32 & 1.5 & 1.7 & 2.1 & 435 & 224 & 103 & 198 & 75 & 41 & 2.2 & 3.0 & 2.5 \\ \hline
			hR & 40 & 44 & 20 & 30 & 33 & 15 & 1.3 & 1.3 & 1.3 & 420 & 230 & 133 & 123 & 49 & 33 & 3.4 & 4.7 & 4.0 \\ \hline
			cP & 42 & 22 & 20 & 21 & 6 & 4 & 2.0 & 3.7 & 5.0 & 187 & 58 & 43 & 45 & 18 & 13 & 4.2 & 3.2 & 3.3 \\ \hline
			cF & 75 & 65 & 48 & 19 & 10 & 6 & 3.9 & 6.5 & 8.0 & 251 & 80 & 43 & 27 & 17 & 7 & 9.3 & 4.7 & 3.9 \\ \hline
			cI & 58 & 14 & 10 & 21 & 6 & 2 & 2.8 & 2.3 & 5.0 & 92 & 15 & 6 & 30 & 5 & 3 & 3.1 & 3.0 & 2.0 \\ \hline
		\end{tabular} }
	\end{table*}

Another interesting observation is that while some stoichiometries are abundant in the oxides 
they are almost absent in the sulfides or the selenides. For example,
there are 299 compounds with the $A_2B_2$O$_7$ stoichiometry (ignoring
order between $M_A$ and $M_B$), but only two $A_2B_2$S$_7$ compounds
and no $A_2B_2$Se$_7$ compounds. Also, there are 71 $A_1B_3$O$_9$
compounds but no $A_1B_3$S$_9$ and $A_1B_3$Se$_9$ compounds. On the
other hand, there are no $A_4B_{11}X_{22}$ oxides, but 20 sulfides and 8 selenides. 
If we require that $M_A>M_B$, there are no oxides of the $A_3B_2X_2$ 
stoichiometry, but 25 sulfides and 7 selenides.

\begin{figure*}
	\centering
  \includegraphics[width=1.0\textwidth,height=0.85\textheight,keepaspectratio]{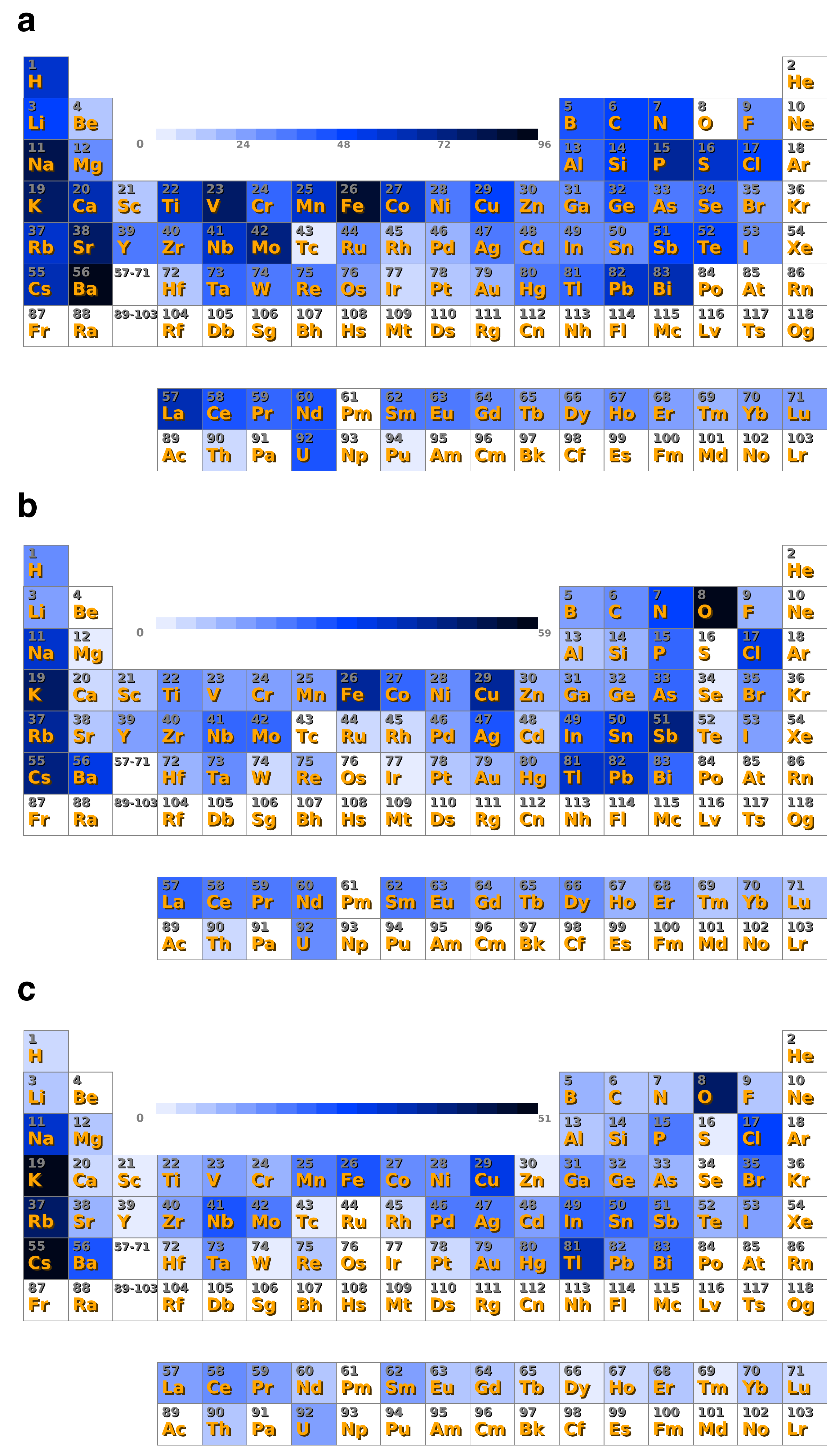}
  \caption{Ternary (\textbf{a}) oxide, (\textbf{b}) sulfide, and (\textbf{c}) selenide stoichimetries {(number of different stoichiometries that include the respective element)
  		. The colors go from no stoichiometries (white) to the maximal number of stoichiometries (dark blue) which is different for each element, 96/59/51 for O/S/Se. High prevalence appears for the alkali metals in all three compound families. An additional island in the transition metals is much more pronounced in the oxides.
  		The sulfides and selenides distributions are nearly identical, and shown high prevalence of oxygen containing ternaries. }
	}
	\label{fig:tern_stoi_periodic_oxide}
\end{figure*}

Again, an important conclusion is that there are many missing stoichiometries, 
Figure \ref{fig:triangles}(d) shows all the possible stoichiometries for $A_xB_yC_z$ for $x,y,z \le 12$, 
clearly showing rich concentration in the middle, which is not the case for oxides, and also to a 
lesser degree to sulfides and selenides. 

\begin{figure*} [t]
	\centering
  \includegraphics[width=1.0\linewidth]{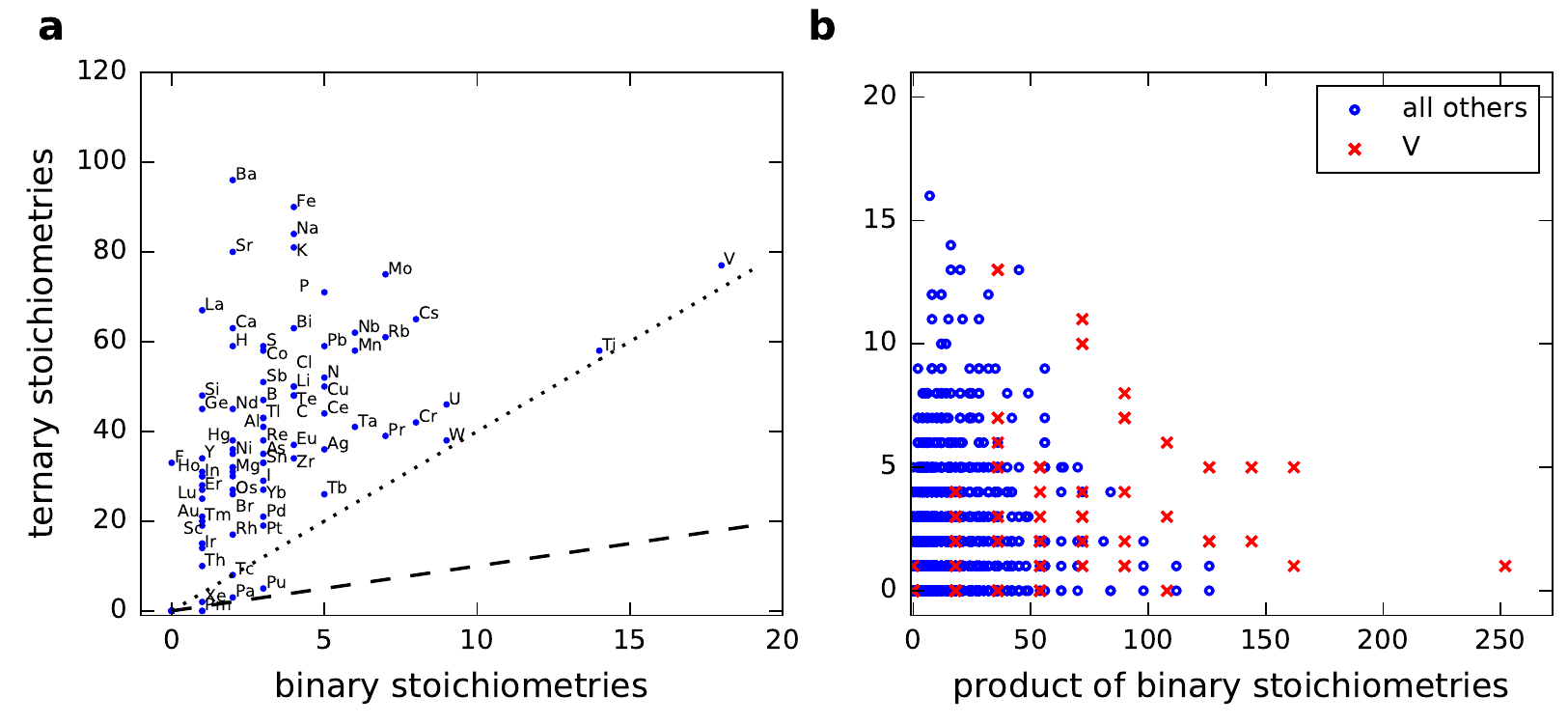}
  \caption{(\textbf{a}) Prevalence of ternary oxide stoichiometries per atom as a function of the prevalence of its binary stoichiometries. 
    The dashed line marks perfect similarity $(y=x)$, and the dotted line marks the ratio $y=4x$.
  (\textbf{b}) Number of oxide ternary stoichiometries as a function of product of participating
  elements numbers of oxide binary stoichiometries. The data for vanadium is shown with red crosses, all the rest is shown with blue circles.
  }
	\label{fig:bintern1a}
\end{figure*}

We can repeat the analysis of the binary stoichiometries and ask how many stoichiometries 
per element are there for the ternaries. 
This is shown in Figure \ref{fig:tern_stoi_periodic_oxide}.
Here, also, the similarity of sulfides and selenides is clear. 
In addition, while there are similarities between the distributions of binary stoichiometries 
per element to the ternary distributions, there are also obvious differences. 
One might guess that there should be a correlation between the binary and ternary distributions. 
This is examined in Figure \ref{fig:bintern1a}(a).

It is evident that the correlation between ternary and binary number of stoichiometries is not strong 
but the minimal number of ternary stoichiometries tends to grow with the number of binary stoichiometries. 
We check this further in Figure \ref{fig:bintern1a}(b), by comparing the number of ternary stoichiometries of 
$A_xB_y$O$_z$ to the product of stoichiometry numbers of $A_x$O$_y$ and $B_x$O$_y$. 
The general trend obtained is an inverse correlation, i.e., as the product of the numbers of binary 
stoichiometries increases, the number of ternaries decreases. This trend can be explained by the following argument: when the two binaries are rich with stable compounds, the ternaries need to compete with more possibilities of 
binary phases, which makes the formation of a stable ternary more difficult. 
In Figure \ref{fig:bintern1a}(b), this trend is highlighted for vanadium, 
the element with the most binary stoichiometries, but this pattern repeats itself for most elements. 
We analyze this behavior for the sulfides and selenides in the SI, similar trends are found but 
they are less pronounced due to a smaller number of known compounds. 

\begin{figure*} [t]
	\centering
	\includegraphics[width=1.0\linewidth]{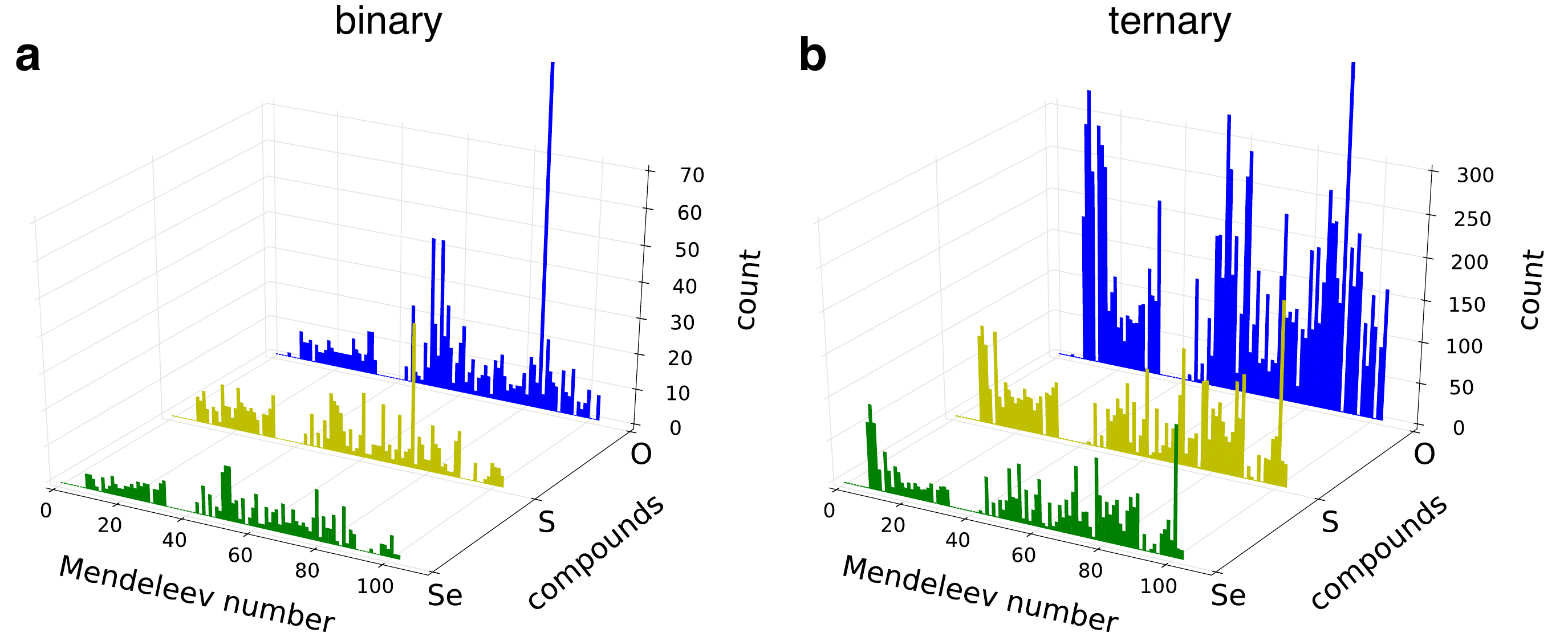}
  \caption{Distribution of the elements in (\textbf{a}) binary and (\textbf{b}) ternary compounds. {The binary oxides exhibit a structures distribution with two prominent peaks. The distributions of the binary sulfides and selenides are less structured and more similar to each other. The distributions of the ternary compounds have higher, wider peaks than
  		their binary counterparts. The relative differences between the oxide, sulfide and selenide
  		distributions remain similar to the distributions of the binaries. }
	}
	\label{fig:mendeleev_distribution_all_in_one}
\end{figure*}

\noindent \textbf{Composition and Mendeleev maps.}
The occurrence of each element in the binary and ternary compound lists has been 
counted and tabulated. 
The results are described in Figure~\ref{fig:mendeleev_distribution_all_in_one}. 
For the binary oxides a very prominent peak appears at $M=85$, the
Mendeleev number of silicon. 
It represents the 185 different silicon oxide
structures types (s.t.) reported in the \ICSD\ database for just a {\it single} stoichiometry, SiO$_2$.
Smaller peaks appear for $M=51$ (titanium, 42 s.t., 14 stoichiometries, 
leading stoichiometry is 
TiO$_2$ with 14 s.t.), $M=54$ (vanadium, 42 s.t.,
18 stoichiometries, leading stoichiometry is VO$_2$ with 10 s.t.), 
$M=56$ (tungsten, 24 s.t., 9 stoichiometries, leading stoichiometry is WO$_3$ with 13 s.t.), 
and $M=45$ 
(uranium, 22 s.t., 9 stoichiometries, leading stoichiometries are UO$_2$ and U$_3$O$_8$ with 6 s.t. each). 
Unlike the silicon peak which is composed of a single stoichiometry, 
the other leading peaks evidently include multiple stoichiometries, reflecting the different chemistry of those elements. 
These differences also carry over into the ternary oxide compounds involving those elements. 
For example, the stoichiometry distribution of silicon oxide ternaries is more tilted towards 
the silicon poor compounds compared to the corresponding distributions of vanadium and titanium ternary oxides, 
as is shown in Figure $S3$ in the SI. 

The distribution of the sulfides is generally much lower than
that of the oxides, due to the much smaller total number of known binaries, but is also more uniformly structured. 
It has one major peak 
for $M=76$ (zinc, 40 s.t., 2 stoichiometries, leading stoichiometry is ZnS with 39 s.t.), 
and quite a few smaller ones such as $M=51$ (titanium, 16 s.t., 5 stoichiometries, leading stoichiometry is TiS$_2$ with 9 s.t.), 
$M=61$ (iron, 18 s.t., 5 stoichiometries, leading stoichiometry is FeS with 6 s.t.), 
$M=67$ (nickel, 16 s.t., 6 stoichiometries, leading stoichiometry is NiS$_2$ with 8 s.t.), 
$M=90$ (phosphorus, 13 s.t., 8 stoichiometries, of which
P$_2$S$_7$, P$_4$S$_9$, P$_4$S$_6$, P$_4$S$_5$ and P$_4$S$_3$ have 2 s.t. each). 
The $M$~=~8--33 region also exhibits a minor concentration of
participating elements. 
The selenides distribution is yet smaller than that of the sulfides, and
even more uniform. 
Several peaks appear, $M=51$ (titanium, 13 s.t., 9 stoichiometries, leading stoichiometry is 
TiSe with 3 s.t.), 
$M=52$ (niobium, 15 s.t., 8 stoichiometries, leading stoichiometry is 
NbSe$_2$ with 8 s.t.), 
$M=53$ (tantalum, 15 s.t., 4 soichiometries, leading stoichiometry is 
TaSe$_2$ with 10 s.t.) and 
$M=79$ (indium, 14 s.t., 5 stoichiometries, leading stoichiometry is In$_2$Se$_3$ with 6 s.t.).
All distributions cover most of the elements except two obvious gaps, one at $M<9$, 
which includes the noble gases and the two heaviest alkali metals, cesium and francium, and another 
at $34\leq M\leq 42$ which represents the heavy actinides. Another gap appears in the sulfide and selenide distributions at $91\leq M\leq 97$, 
which reflects the rarity of polonium and astatine compounds and shows that the elements of the 6A column, 
except oxygen, do not coexist, in the known compounds, with each other
or with the heavier halogen iodine.

The element occurance distributions for the oxide, sulfide and
selenide ternaries exhibit greater similarity than the
corresponding binary distributions. The most apparent difference, however, is the
most common component, which is sulfur, $M=90$, in the
oxides, but oxygen itself, $M=101$, in the sulfides and selenides. The
sulfide and selenide distributions are almost the same, except for
generally lower numbers in the selenides (due to the smaller total
number of compounds) and an apparent lower participation of the
lantanides $M$~=~17--35.
 
\begin{figure*}
	\centering
	\includegraphics[width=1.0\textwidth,height=0.85\textheight,keepaspectratio]{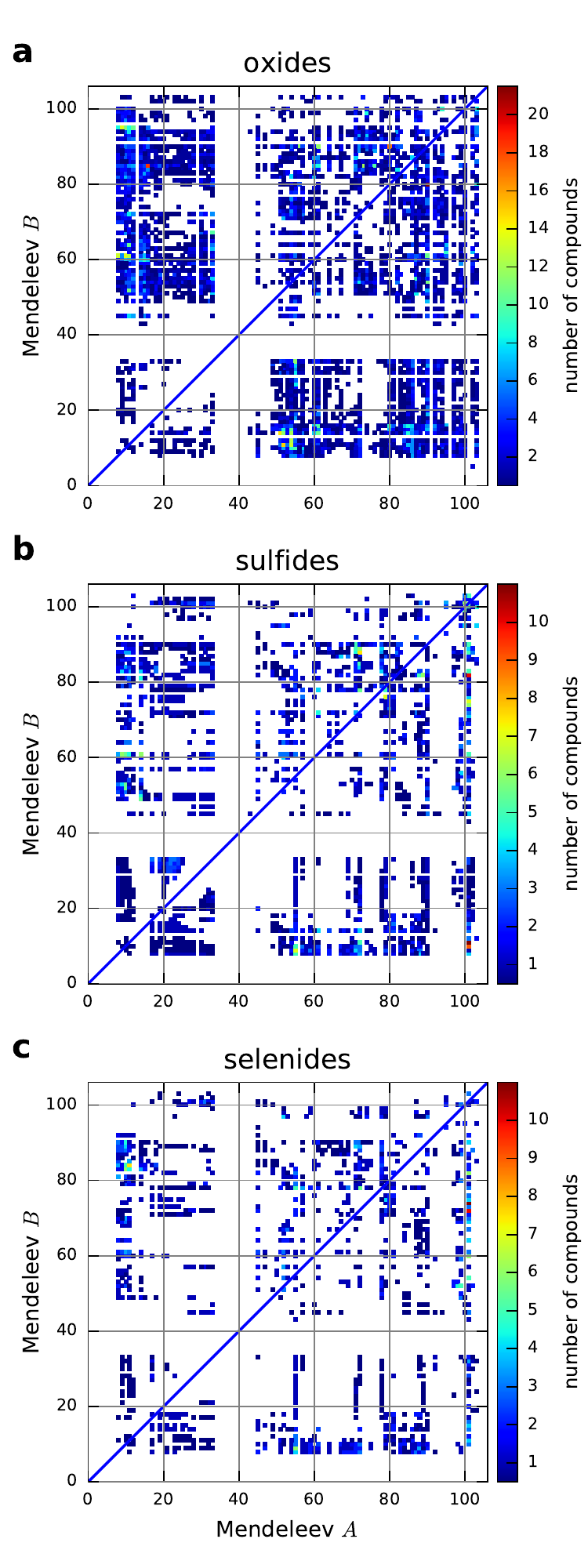}
	\caption{Mendeleev maps of ternary (\textbf{a}) oxide $A_xB_y$O$_z$, (\textbf{b}) sulfide $A_xB_y$S$_z$
		and (\textbf{c}) selenide $A_xB_y$Se$_z$ compounds. 
		It is assumed that $x\geq y$ with the $x$-axis indicating $M_A$
		and the $y$-axis $M_B$. 
		If the stoichiometry is such that $x=y$, the compound is counted as $0.5 A_xB_y$O$_z + 0.5 B_xA_y$O$_z$. {A color scheme is used to represent the compound count for each composition, blue means the minimal number (one) and green means the maximal number which is different for each element.}
	}
	\label{fig:mendeleev_bigger_x_upper_all_in_one}
\end{figure*}

\begin{figure*} 
	\includegraphics[width=\textwidth]{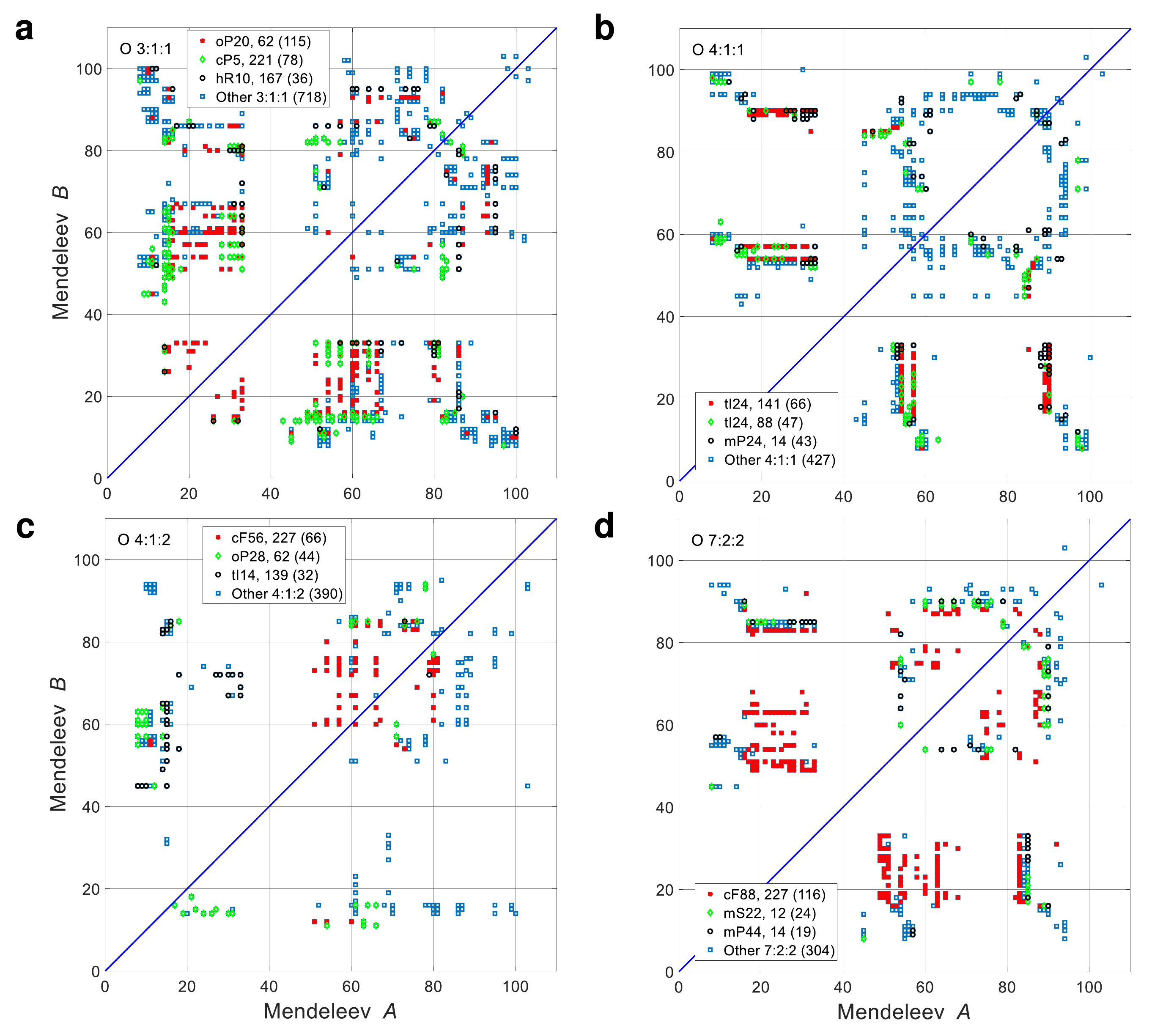}
	\caption{Three leading structure types in each of the four leading stoichiometries in oxide ternaries:
		(\textbf{a}) $A_1B_1$O$_3$,
		(\textbf{b}) $A_1B_1$O$_4$,
		(\textbf{c}) $A_1B_2$O$_4$, and
		(\textbf{d}) $A_2B_2$O$_7$. 
		The legend box appears at a region with no data points. The number in parenthesis is the number of compounds for this structure type, for ``Other", it refers to the total number of compounds with this stoichiometry.
	}
	\label{fig:mendeleev_oxide_prototypes}
\end{figure*}

\begin{figure*} 
	\includegraphics[width=\textwidth]{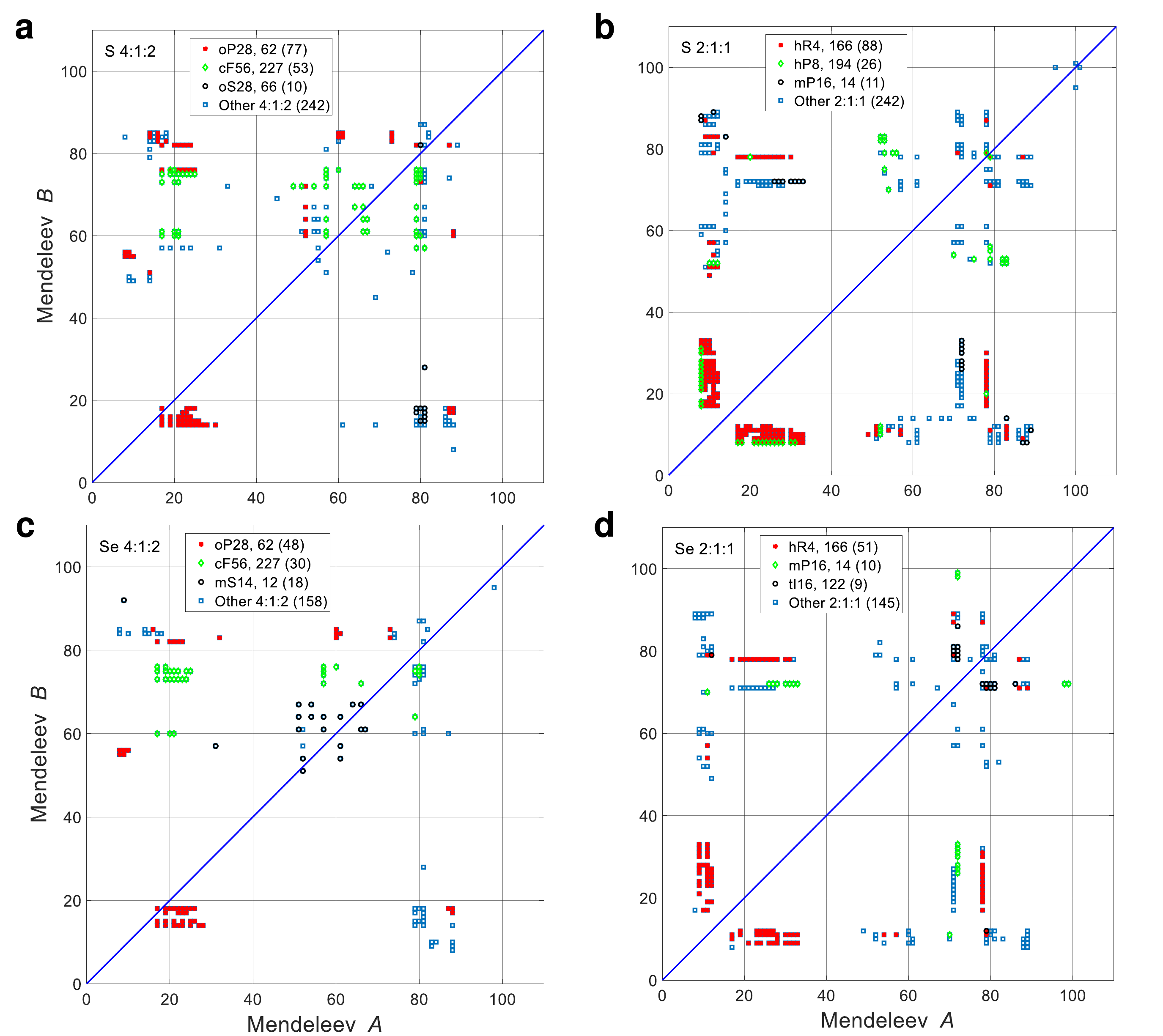}
	\caption{Three leading structure types in each of the two leading stoichiometries in sulfur and selenium ternaries:
		(\textbf{a}) $A_1B_2$S$_4$,
		(\textbf{b}) $A_1B_1$S$_2$,
		(\textbf{c}) $A_1B_2$Se$_4$, and
		(\textbf{d}) $A_1B_1$Se$_2$. The number in parenthesis is the number of compounds for this structure type, for ``Other", it refers to the total number of compounds with this stoichiometry.
	}
	\label{fig:mendeleev_sulfur_selenium_prototypes}
\end{figure*}

Mendeleev maps for the ternaries are shown in 
Figures~\ref{fig:mendeleev_bigger_x_upper_all_in_one}-\ref{fig:mendeleev_sulfur_selenium_prototypes}.
Figure~\ref{fig:mendeleev_bigger_x_upper_all_in_one} shows the cumulated maps for all
stoichiometries reported for the respective ternary family.
They reflect the same major gaps as the binary distributions.
The maps show that most of the reported compositions are represented by one or two compounds
with just a few hotspots that include up to 20 compounds in the oxides and
10 compounds in the sulfides and selenides.
The oxides map is obviously denser, reflecting the much richer, currently known, chemistry of the oxides compared
to the other two elements.
The chemistry becomes more constrained as we proceed down the periodic table column from
oxygen to sulfur and then to selenium.

Next, we examine maps of specific stoichiometries.
Maps of a few notable oxide stoichiometries and their 
leading structure types are shown in Figure~\ref{fig:mendeleev_oxide_prototypes}. 
These maps reflect the dominant features of the 
full oxide ternaries map (Figure~\ref{fig:mendeleev_bigger_x_upper_all_in_one}),
but with significant new additional gaps of absent compounds. These gaps are naturally
wider for less prevalent stoichiometries, i.e., the map of the most
prevalent stoichiometry,  $A_1B_1$O$_3$, is denser than the three
other maps in  Figure~\ref{fig:mendeleev_oxide_prototypes}.
Different structure types in all stoichiometries tend to accumulate at
well defined regions of the map. The separation between them
is not perfect, but is similar to that exhibited by the classical Pettifor maps for
binary structure types \cite{pettifor:1984,pettifor:1986}. 
A similar picture is obtained for the sulfide and selenide structure types, although more sparse
(Figure~\ref{fig:mendeleev_sulfur_selenium_prototypes}).
It is interesting to note that the maps of, e.g., 
$A_1B_2C_4$ ($C=$ O, S, Se), show similar patterns in the map for oxides 
(Figure \ref{fig:mendeleev_oxide_prototypes}) and sulfides/selenides 
(Figure \ref{fig:mendeleev_sulfur_selenium_prototypes}) --- 
suggesting that similar elements tend to form this stoichiometry. 
In the same manner, the 2:1:1 stoichiometry shows very similar patterns in oxides, sulfides and selenides (see also Figure $S6$ in the SI). 

\begin{figure*} 
\includegraphics[width=\linewidth]{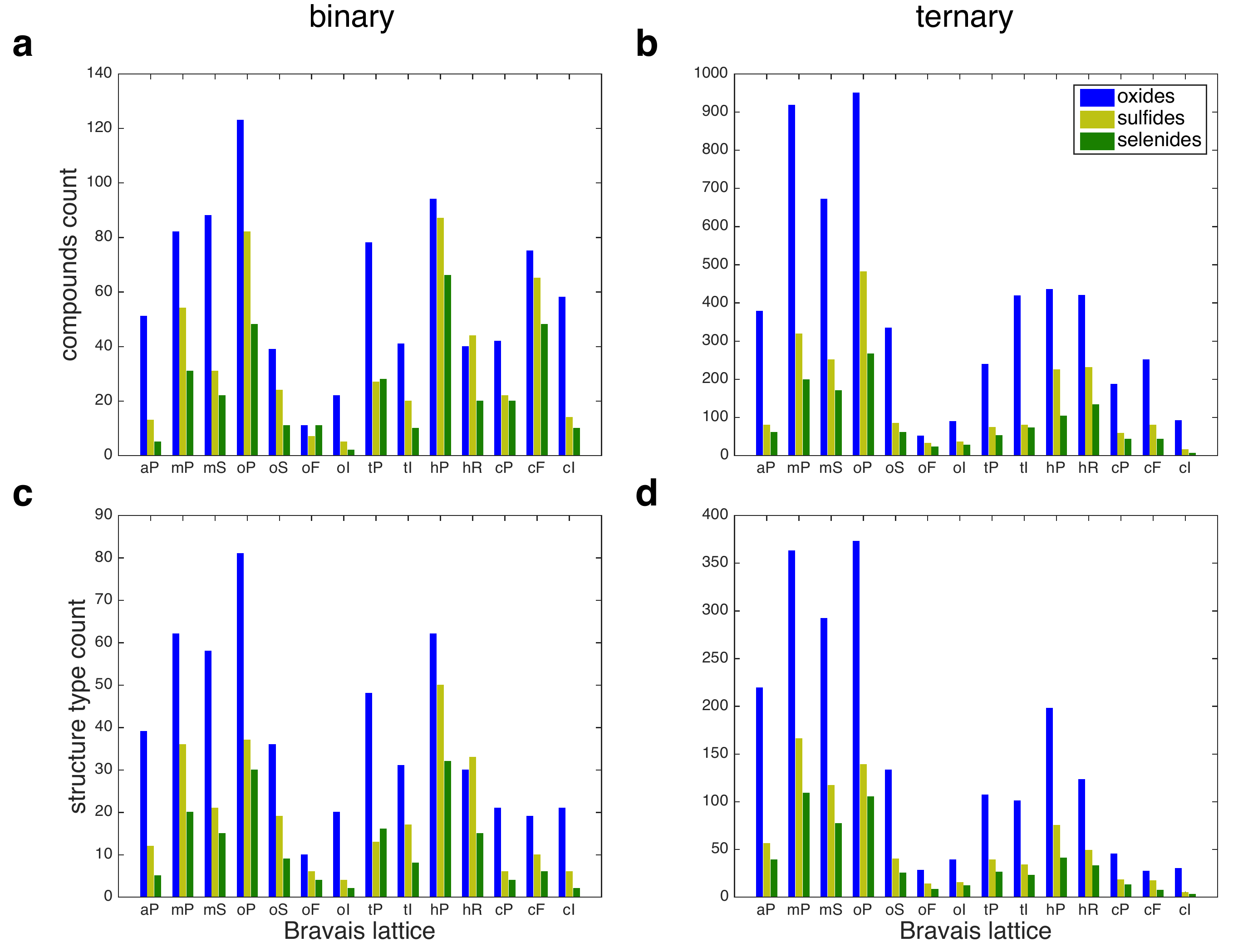}
\caption{Number of compounds (\textbf{a} and \textbf{b}) and
structure types (\textbf{c} and \textbf{d}) for each Bravais lattice.
Binaries are on the left (\textbf{a} and \textbf{c}) and
ternaries on the right (\textbf{b} and \textbf{d}). {Oxides are shown in blue, sulfides in light green and selenides in darker green. All six  distributions (binary and
	ternary oxides, sulfides and selenides) are double peaked with a
	local minimum for the orthorhombic face and body centered lattices. The high symmetry
	cubic lattices are also relatively rare. This reflects the complex spatial arrangement of
	the compound forming electrons of the 6A elements, which does not favor the
	high symmetry of these
	structures.}
}
\label{fig:bravais_combined}
\end{figure*}

\noindent \textbf{Symmetries.}
The distribution of the compounds and structure types among the 14 Bravais lattices 
is presented in Table~\ref{table:bravais_lattice_distribution} and 
Figure~\ref{fig:bravais_combined}. 
It is interesting to note that in all six cases (binary and ternary oxides, sulfides and selenides) 
the distribution is double peaked, with the majority of the compounds belonging to the 
monoclinic and orthorhombic primitive lattices, 
and a smaller local maximum at the hexagonal and tetragonal lattices. 
All distributions exhibit a local minimum for the orthorhombic face and body centered lattices. 
The high symmetry cubic lattices are also relatively rare. 
This reflects the complex spatial arrangement of the compound forming electrons of oxygen, sulfur and selenium, 
which does not favor the high symmetry cubic structures or the 
densely packed face and body centered orthorhombic structures.

\begin{figure*} [t]
\centering
\includegraphics[width=1.0\linewidth]{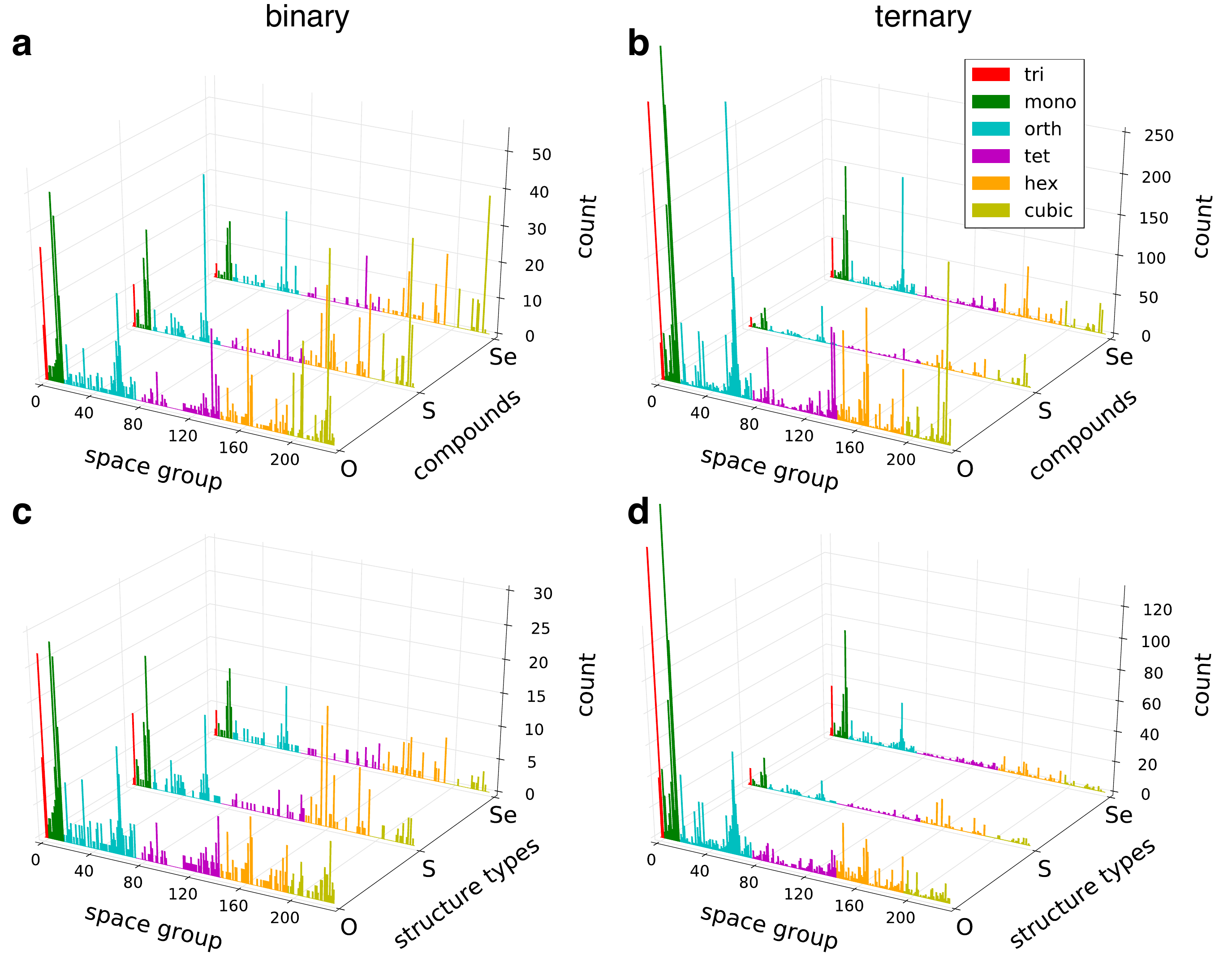}
\caption{Distribution of compounds and structure types among the 230 space groups.
Binaries are on the left (\textbf{a} and \textbf{c}) and
ternaries on the right (\textbf{b} and \textbf{d}).
Compounds are depicted on the top (\textbf{a} and \textbf{b})
and structure types on the bottom (\textbf{c} and \textbf{d}).
}
\label{fig:symmetry_distribution_of_structures}
\end{figure*}

Figure~\ref{fig:symmetry_distribution_of_structures} 
shows a more detailed distribution of the compounds among the different space groups. 
The binary compounds show a distinct seesaw structure, with a few 
local peaks near the highest symmetry groups of each crystal system. 
The corresponding ternary distributions have three sharp peaks in the triclinic, 
monoclinic and orthorhombic systems, and much smaller peaks in the hexagonal and cubic groups. 
It is interesting to note that the three compound families, exhibit distributions of very similar structure. 
The oxide distributions are the densest, simply due to the existence of more oxide compounds in the database, 
and become sparser in the sulfide and selenide cases. 
The compounds of all these families are distributed among a rather limited number of space groups, 
with most space groups represented by just a single compound or not at all.

\begin{figure*}
\centering
\includegraphics[width=1.0\linewidth]{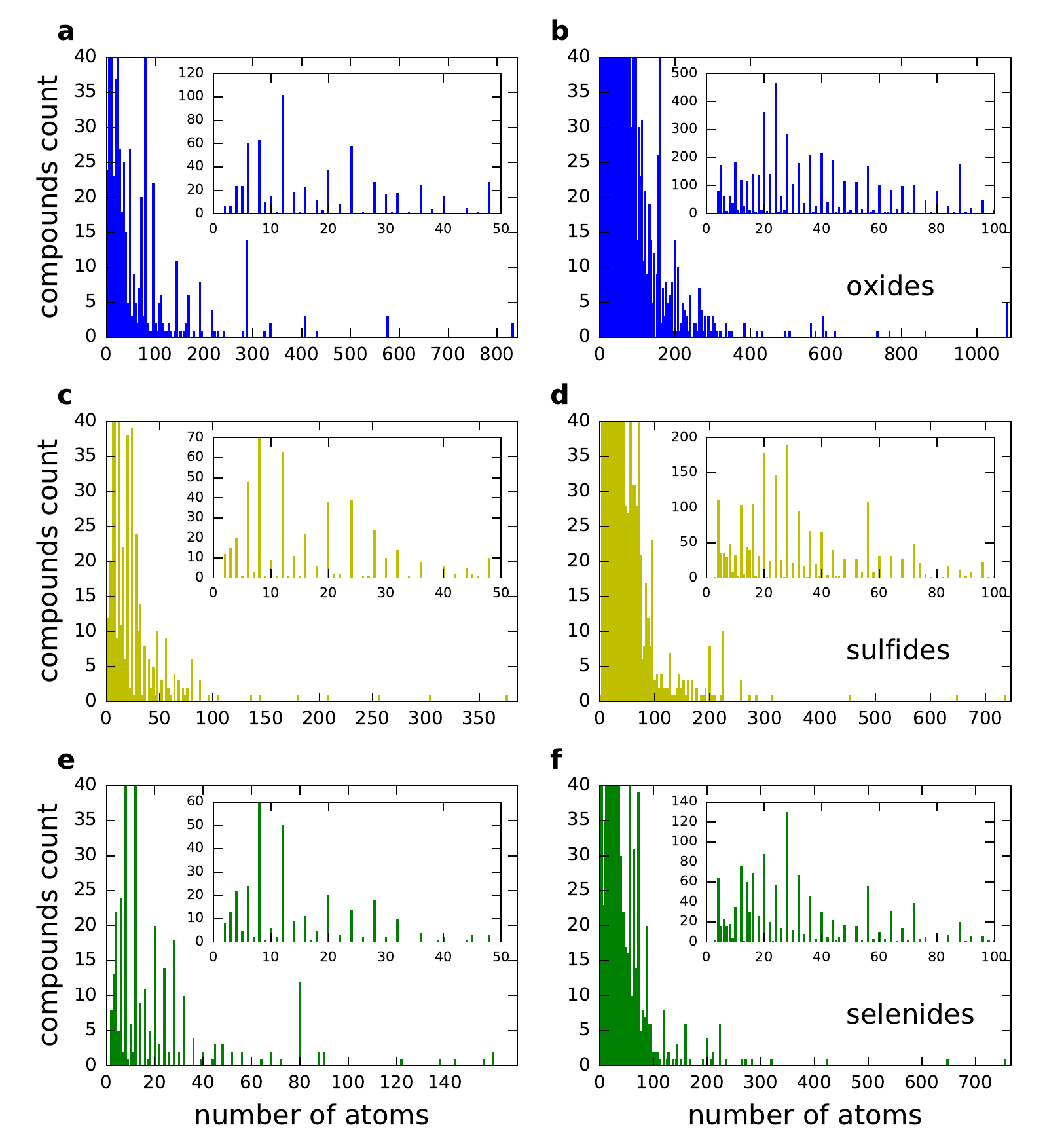}
\caption{Unit cell size distribution for all compounds. 
Binaries are on the left (\textbf{a}, \textbf{c} and \textbf{e}) and
ternaries on the right (\textbf{b}, \textbf{d} and \textbf{f}).
Oxides are at the top (\textbf{a} and \textbf{b}), 
sulfides in the middle (\textbf{c} and \textbf{d}) and
selenides at the bottom (\textbf{e} and \textbf{f}).
The insets show the compounds with up to 50 atoms per unit cell in each case. 
{All distributions exhibit long tails of rare very large unit cells which extend much further in the oxides. The dense cores of the distributions reflects the higher prevalence of oxides and are very similar for the sulfides and selenides.}}
\label{fig:number_of_atoms_distribution}
\end{figure*}

\noindent \textbf{Unit cell size.}
The distributions of unit cell sizes (i.e., the number of atoms per unit cell) for the six compound families we discuss
are shown in Figure~\ref{fig:number_of_atoms_distribution}. 
All of these distributions have strong dense peaks at small cell sizes and decay sharply at sizes above a few tens of atoms. 
However, the details of the distributions differ quite significantly from group to group. 
Among the binaries, the oxides exhibit the highest and widest peak with
its maximum of 102 oxide binary compounds located at 12 atoms per cell. 
90\% of the oxide binaries have less than 108 atoms in the unit cell and 50\% of them have less than 24 atoms.
The sulfides distribution has a lower and narrower peak of 70 compounds at 8 atoms. 
The distribution of the selenides has a still lower peak of 60 compounds at 8 atoms. {The fact that oxygen has a peak at 12 atoms in the unit cell and not at 8 as the sulfides and selenides, is related to the fact that binary oxides prefer the $A$O$_2$ stoichiometry over $A$O, where as both sulfides and selenides prefer the 1:1 stoichiometry over 1:2. This is probably related to the different chemistry of oxygen vs. sulfur and selenium. Additional computational analysis would be required to fully understand the effect of the different chemistry on the stoichiometry and number of atoms.}
Detailed data for these dense parts of the distributions is tabulated in Table $S12$ the Supporting Information (SI). 
The oxides distribution exhibits the longest tail of the binaries, with the largest 
binary oxide unit cell including 576 atoms. 
The largest binary sulfide and selenide unit cells include 376 and 160 atoms, respectively. 

The distributions of the ternary compounds have higher, wider peaks and longer tails than 
their binary counterparts.
The relative differences between the oxide, sulfide and selenide distributions 
remain similar to the distributions of the binaries. 
The oxide ternaries exhibit a high and wide peak. 
Its maximum of 465 compounds is located at 24 atoms per cell, and
90\% of the compounds have less than 92 
atoms in the unit cell and 50\% of the compounds have less than 32 atoms.
As in the binary case, the distribution of the ternary sulfides has a lower and narrower peak than the oxides, 
where the maximum of 190 compounds at 28 atoms and 90\% of the compounds have less than 72 atoms in the unit cell. 
The distribution of the selenides has a still lower and narrower peak, where the corresponding numbers are 
130 compounds at 28 atoms and 90\% of the compounds having less than 28 atoms in the unit cell.
Detailed data for these dense parts of the distributions is shown in Table $S13$ of the SI.
The ternary oxides distribution exhibits the longest tail of 
the three types, with the largest oxide ternary unit cell having 1,080 atoms. 
The largest ternary sulfide and selenide unit cells have 736 and 756
atoms, respectively.

It should be noted that large unit cells, within the tails of all distributions, tend to have very few representatives,
with just one compound with a given unit cell size in most cases. 
Notable exceptions are local peaks near $80$ atoms per unit cell in the binary 
distributions and near 200 atoms per unit cell in the ternary distributions. 
The oxide distributions exhibit additional peaks, near 300 atoms per unit cell for the 
binaries and near 600 atoms per unit cell for the ternaries. 
These minor peaks may indicate preferable arrangements of cluster-based structures. 

\section*{Summary}
We present a comprehensive analysis of the statistics of the binary
and ternary compounds of oxygen, sulfur and selenium. This analysis and the visualization tools presented here are 
valuable to finding trends as well as exceptions and peculiar phenomena.

Oxygen has a higher electronegativity (3.44) than sulfur
(2.58) and selenium (2.55), which are similar to each other. 
Therefore, one can expect that oxygen will form compounds with a stronger ionic character. 
{Oxygen is 1000 times more abundant than sulfur, and more than $10^6$ times than selenium\cite{wedepohl1995composition}, however, it has less than two times the number of binary compounds compared to sulfur and $2.5$ that of selenium. Hence, the abundance of those elements plays a little role in the relative numbers of their known compounds  }. These important differences are reflected in our analysis by 
the significantly larger fraction of oxygen rich compounds compared to
those that are sulfur or selenium rich. 
Structure type classification also shows that there is little overlap between the oxygen 
structure types to sulfur or selenium structure types, while 
sulfur and selenium present a much higher overlap. The gaps in these overlaps, especially between 
the sulfides and selenides, indicate that favorable candidates for new compounds 
may be obtained by simple element substitution in the corresponding structures.
In particular, structures than are significantly more common in one family, 
such as KrF$_{2}$ in the oxides, may be good candidates for new compounds in another. 
Comparison of these three 6A elements binary and ternary
compounds shows significant differences but also some similarities in the symmetry
distributions among the various Bravais lattices and their
corresponding space groups. In particular, the majority of structure types in all three families have a few or single
compound realizations. This prevalence of unique structure types suggests a ripe field for identification of currently unknown compounds, 
by substitution of elements of similar chemical characteristics.
In addition, the analysis of the distribution of known compounds among symmetry space groups and, 
in particular, their apparent concentration in specific hot spots of this 
symmetry space may be serve as a useful insight for searches of potential new compounds. 

An important observation is the existence of different gaps (missing stoichiometries) in the stoichiometry distribution of the oxide 
binary compounds compared to the sulfides and selenides (Figures 2 and 3, and Table $S7$ in the SI).
Stoichiometries such as 5:7 appear in the oxides but are missing in the sulfides and selenides. More rare are non-overlapping gaps between the selenides and sulfides, e.g. 6:1 and 5:7. These should be prime candidates for new compounds by element substitution between the two families. {Future work would be directed at exploiting these discrepancies to search for new compounds within different subsets of those compound families.}

Specific elements tend to present very different stoichiometry distributions, for example, silicon forms only 
one oxide stoichiometry (SiO$_2$) while transition metals such as titanium and vanadium present 
14 and 18 different stoichiometries respectively. 
These differences clearly reflect the different chemistry of those elements, while the large number of reported 
SiO$_2$ structures might reflect research bias into silicon compounds.

Another important finding is that there is an inverse correlation between the number of ternary stoichiometries 
to the product of binary stoichiometries of participating elements. 
This can be caused by the fact that there are too many binary phases and hence it becomes 
difficult to create a stable ternary that competes with all of them.

A Mendeleev analysis of the common structure types of these
families shows accumulation of different structures at
well defined regions of their respective maps, similar to the well-known Pettifor maps of binary structure types. 
Furthermore, at least for some of the stoichiometries, similarity of the maps for a 
given stoichiometry is demonstrated across all three elements. 
These maps should therefore prove useful for predictive purposes regarding the existence
of yet unknown compounds of the corresponding structure types. 
{Future work will be directed at exploiting identified non-overlapping gaps in the 
Mendeleev maps for a directed search of new compounds in these families. 
Complementary properties (e.g. partial charges, bond analysis, electronic properties) 
should be incorporated in the analysis to reveal additional insights of the aforementioned trends among the three elements.}

\section*{Acknowledgements}
AN acknowledges
financial support from the Israeli National Nanotechnology Initiative (INNI, FTA project).  
AN and OL acknowledge financial support from the Pazi foundation.                           
CO acknowledges support from the National Science Foundation Graduate 
Research Fellowship under Grant No. DGF1106401.
SC and CO acknowledge support by DOD-ONR (N00014-17-1-2090) and DOE (DE-AC02-05CH11231),    
specifically the Basic Energy Sciences program under Grant \# EDCBEE.                       
SC acknowledges support by the Alexander von Humboldt-Foundation.                           
The authors thank Drs. Cormac Toher, David Hicks, Natalio Mingo, Luca Ghiringhelli, and                  
Jes\'{u}s Carrete for various technical discussions.                                        

\newcommand{\Ozolins}{Ozoli\c{n}\v{s}}

\end{document}